# *Chandra* Characterization of X-ray Emission in the Young F-Star Binary System HD 113766

## C.M. Lisse[1], D.J. Christian[2], S.J. Wolk[3], H.M. Günther[4], C.H. Chen[5], C.A. Grady[6]

[1]Planetary Exploration Group, Space Department, Johns Hopkins University Applied Physics Laboratory, 11100 Johns Hopkins Rd, Laurel, MD 20723   carey.lisse@jhuapl.edu

[2]Department of Physics and Astronomy, California State University Northridge, 18111 Nordhoff Street, Northridge, CA 91330 damian.christian@csun.edu

[3]*Chandra* X-ray Center, Harvard-Smithsonian Center for Astrophysics, 60 Garden Street, Cambridge, MA, 02138  swolk@cfa.harvard.edu

[4]Massachusetts Institute of Technology, Kavli Institute for Astrophysics and Space Research, 77 Massachusetts Avenue, NE83-569, Cambridge, MA 02139   hgunther@mit.edu

[5]STScI, 3700 San Martin Drive, Baltimore, MD 21218 cchen@stsci.edu

[5]Eureka Scientific and Goddard Space Flight Center, Code 667, NASA-GSFC, Greenbelt, MD 20771 carol.a.grady@nasa.gov

23 Pages, 3 Figures

Proposed Running Title: **_Chandra_ Observations of HD 113766 X-ray Emission**

Please address all future correspondence, reviews, proofs, etc. to :

Dr. Carey M. Lisse

Planetary Exploration Group, Space Department

Johns Hopkins University, Applied Physics Laboratory

11100 Johns Hopkins Rd

Laurel, MD 20723

240-228-0535 (office) / 240-228-8939 (fax)

Carey.Lisse@jhuapl.edu

## Abstract

Using *Chandra* we have obtained imaging X-ray spectroscopy of the 10-16 Myr old F-star binary HD 113766. We individually resolve the $1.4''$ separation binary components for the first time in the X-ray and find a total 0.3–2.0 keV luminosity of $2.2 \times 10^{29}$ erg/sec, consistent with previous RASS estimates. We find emission from the easternmost, infrared-bright, dusty member HD 113766A to be only ~10% that of the western, infrared-faint member HD 113766B. There is no evidence for a 3rd late-type stellar or sub-stellar member of HD113766 with $L_x > 6 \times 10^{25}$ erg s$^{-1}$ within 2' of the binary pair. The ratio of the two stars' X-ray luminosity is consistent with their assignments as F2V and F6V by Pecaut *et al.* (2012). The emission is soft for both stars, $kT_{Apec} = 0.30$ to 0.50 keV, suggesting X-rays produced by stellar rotation and/or convection in young dynamos, but not accretion or outflow shocks which we rule out. A possible $2.8 \pm 0.15$ ($2\sigma$) hr modulation in the HD 113766B X-ray emission is seen, but at very low confidence and of unknown provenance. Stellar wind drag models corresponding to $L_x \sim 2 \times 10^{29}$ erg s$^{-1}$ argue for a 1 mm dust particle lifetime around HD 113766B of only ~90,0000 years, suggesting that dust around HD 113766B is quickly removed, whereas dust around HD 113766A can survive for > $1.5 \times 10^6$ yrs. At $10^{28}$–$10^{29}$ erg s$^{-1}$ luminosity, astrobiologically important effects, like dust warming and X-ray photolytic organic synthesis, are likely for any circumstellar material in the HD 113766 systems.

# 1. Introduction.

We report here on an analysis of *Chandra* soft X-ray observations of HD 113766, a young (10-16 Myr old, Mamajek *et al.* 2002; Chen *et al.* 2006, 2011; Lisse *et al.* 2008; Pecout *et al.* 2012), F-star binary stellar system of near-solar metallicity (Fe/H = -0.1, Nordström *et al.* 2004), located at a distance of 123 $^{+18}/_{-14}$ pc (8.16 mas *Hipparcos* parallax) from the Earth (van Leeuwen 2007). Little is known about this system in the X-ray, other than it is a reported unresolved RASS source of luminosity 2.1 +/- 0.7 x $10^{29}$ erg $s^{-1}$. On the other hand, in the optical/IR, the system is very interesting. With two component stars of nearly identical age characterized by F spectral types in the Sco-Cen star-forming association, attention had been called to this system since its association with object IRAS 13037–4545 in the IRAS Point Source Catalogue (Backman & Paresce 1993). More recent work by Meyer *et al.* (2001), Lisse *et al.* (2008), and Chen *et al.* (2005, 2006, 2011) have confirmed that the system exhibits unobscured photospheres and that HD 113766A exhibits one of the largest IR flux excesses measured ($L_{IR}/L_* = 0.015$), with no detectable $H_2$ emission. HD 113766A thus belongs to the class of post-T Tauri objects characterized by young ages of 5–30 Myr, no leftover primordial gas, and large quantities of excess mid-IR emission from circumstellar dust. On the other hand, there is no evidence for circumstellar dust orbiting the companion, a coeval F-star HD 113766B (Meyer *et al.* 2001), and studies of young stellar clusters (e.g. h and χ Persei, Currie *et al.* 2007, 2008) would have led us to expect a few similar dust forming collisions every Myr in HD 113766B.

Late F stars are typically strong X-ray emitters at young age. In fact, the X-ray luminosity function (XLF) of the Hyades open cluster (age 600 Myr) peaks at late F stars (Stern *et al.* 1995). While X-ray emission of O to mid B-type stars is attributed to dissipating shocks in radiation driven winds, and low-mass GKM stars have strong convection leading to an an αω or α$^2$ dynamo, the origin of emission in late B to early F stars is not so clear. While such stars have a convection zone, the ratio of the X-ray to bolometric flux is much smaller than in typical PMS stars. E.g., Collins *et al.* (2009) find the 10 Myr debris disk host HD 100453 (A9Ve) to have log $L_X/L_{bol} \sim -5.9$, whereas the typical value for PMS GKM stars is about -3.5 (Feigelson *et al.* 2005). Intermediate mass stars also appear to have softer spectra than their lower mass brethren. For example, HD 100453 and the similarly aged 51 Eri (F0V) have coronal temperatures of about 0.2 keV as opposed to 1-2 keV for similarly aged GKM stars (Collins *et al.* 2009, Feigelson *et al.* 2006). Finally, the "FIP-effect" in which elements with first ionization potential below about 10 eV are observed to be enhanced in abundance by a factor of about 3 in the solar corona (Draker et al. 1995), appears to be absent in some F stars such as τ Boo A (an F7V; Maggio *et al.* 2011), and Procyon (an F5IV; Raassen *et al.* 2002 - but see Wood & Laming 2013 for a counter example).

The twin F stars in the HD 113766 system are thus an interesting and useful couple to study. Close enough (separated by only 1.4″, or 170 AU) to be in the same ISM environment, yet far enough separated that they influence each other's circumstellar environment within 100 AU minimally, and formed at the same time with about the same total mass, they are a natural testbed for trying to understand the mechanisms of exosystem formation. As Myr-old F-stars can be expected to be fast rotators, and more convective and X-ray active than their main sequence cousins, we would have naively expected both stars to be rapidly rotating, highly convective, and X-ray bright. We thus obtained *Chandra* observations of HD 113766 because: (a) the system contained a well-known and well-studied IRAS and WISE debris disk while also being a known RASS source, and therefore observable by *Chandra,* a rare combination; (b) using *Chandra*, we could produce the very first resolved maps for the 1.4″ wide binary; and (c) it was important to measure the first resolved X-ray spectrum of the system in order to characterize the stellar wind environment of each of its stars, and understand their effect on the dense dust and ice belts around HD 113766A.

## 2. Observations.

In this Section we present the circumstances and results of our 2010 *Chandra* observations of the HD 113766 system. In the next Section, §3, we will discuss their implications.

## 2.1 CXO Photometry & Luminosity.

We observed the HD 113766 binary using *Chandra* ACIS-S spectroscopy under CXO program OBSID 12384 on 02 Dec 2010 UT. For the observation, no filters or gratings were used, and the stars were centered in the sweet spot of the S3 chip. We re-processed the *Chandra* data with CIAO version 4.8 (Fruscione *et al.* 2006), which applied the energy dependent sub-pixel event repositioning. The total program observing time was 39 ksec, and the total on-target observing time was 37.3 ksec, in which 1509 total raw photons were detected using a source extraction radius of 20″ and an energy filter in the range 0.3-2.0 keV. The extraction region was centered halfway between the two resolved sources' centers. The background was estimated from a large, source-free region on the same chip to be 0.0120 counts pixel$^{-1}$ s$^{-1}$ in 37.2 ksec.

After subtracting the background we determine a total of 1370 source counts, and a source count rate of 0.0366 cps. Using the RASS PSPC rate of 0.034 ± 0.014 cps (ROSAT All-Sky Survey, Voges *et al.* 1999) and assuming a Raymond-Smith coronal plasma model with log T = 6.2 and solar abundances (Raymond & Smith 1977), W3PIMMS indicated a total HD 113766A+B effective *Chandra* count rate of 0.032 cps (assuming a ¼ sub-array to address pile-up). Thus the observed average *Chandra* count rate was within 10% of the RASS count rate extrapolated to the ACIS-S (assuming coronal emission).

The photometry time series (light curve) of the *Chandra* observations is shown for the 2 sources in Figure 1. The 10:1 relative level of brightness of the two sources remains stable over the l0.5 hrs of *Chandra* observation to within the statistical photon noise. There are qualitatively, however, potential variations in the HD 113766B lightcurve that could be periodic. Analyzing the data for possible sinusoidal variations, we find a number of possible solutions with periods ranging from 2.61 to 2.92 hrs, with a best fit solution at 2.87 hrs, a peak-to-peak amplitude of 20%, and $\chi^2_\nu$ = 0.98 for 36 degrees of freedom (dof). However we also find a null periodic solution with $\chi^2_\nu$ = 1.23, and note that the 95% confidence limit of the $\chi^2_\nu$ distribution for 36 dof is 1.42. While it is tempting to assign this periodicity to rotationally induced variability, the equatorial velocity implied for a 1.35 R$_{Sun}$ F6V star is 423 km s$^{-1}$, higher than the predicted breakup speed for a solar abundance

F-star (for $M_* = 1.2$ - $1.6$ $M_{Sun}$, $v_{breakup} \sim 300$ km s$^{-1}$). The implied equtorial velocity is also approximately a factor of 2 higher than the fastest known stellar rotators (Glebocki & Gnaciński 2003, Chen *et al.* 2011). On the other hand, Chen *et al.* (2011) have listed a value of vsini = 93 km s$^{-1}$ for HD113766B, and Smith *et al.* (2012) and Olafsson *et al.* 2013 have published models of the HD113766A disk with inclination i < 10°; if HD113766B has a similar inclination, it could have an equatorial rotation velocity on the order of 400 km s$^{-1}$ consistent with the observations. We thus note the ***possible*** x-ray periodicity in our HD113766B data for future reference, but also note that we do not understand its source if it is real. It will take a deeper, longer set of X-ray observations than is presented here to robustly verify this possibility.

## 2.2   CXO Imaging.

On the S3 chip, we detected 2 closely spaced X-ray sources in the raw HD 113766 imagery (Fig 2a). Image deconvolution was able to better separate these down to the 0.2″ scale, clearly showing two separate sources (Fig 2b). The centers of these lie ~1.4″ apart, and are entirely consistent with the optical and IR locations and separation of ~1.4″ for the two stars (Lisse *et al.* 2008 and references therein), with the optical and IR-brighter HD 113766A to the East, and the optical and IR-fainter but X-ray brighter HD 113766B to the West. The observed flux is very asymmetrically distributed, with ~90% arising from the optically fainter HD 113766B Western binary member, and ~10% arising from the optically brighter HD 113766A Eastern binary member.

In the 2' x 2' subarray field around the HD113766 binary pair, we performed a companion search for potential nearby X-ray sources in the field. There is no evidence in our data for any separate, x-ray bright 3rd member of HD113766 within the 2' of the binary pair with flux greater than 3 times the image background level. Given the background level of ~1.1 x 10$^{-7}$ cps per pixel, and an ACIS-S 90% encircled energy radius of 5 pixels, this implies a 3σ upper limit for any X-ray object in the field of 8.8 x 10$^{-6}$ cps (approximately equivalent to L$_x$ ~ 6 x 10$^{25}$ erg s$^{-1}$, assuming kT = 0.40 keV) for an object at the 130 pc distance of HD 113766A/B. This null result is consistent with the lack of any RASS sources in a 4.2′ x 4.2′ field centered on HD 113766A/B other than the binary system itself (Voges *et al.* 1999).

Any optically faint, coeval 10-16 Myr old KM stellar or L/T brown dwarf class object should have been easily detected in the our ACIS-S imaging (see, for example, the strong *Chandra* detection of the M2.5V HR4796B binary companion to the A0V HR 4796A disk system by Drake *et al.* 2014). Further, the x-ray spectra we report below for the 2 stars is much too soft to be produced by an x-ray active stellar object (Collins *et al.* 2009). While the Chandra data cannot rule out very close-in, optically faint substellar companions within 0.5" of the HD113766 stars that may be x-ray active, no substantial brown-dwarf like infrared excess above the stellar photospheres of either component A or B has been found (Lisse *et al.* 2008, Olafsson *et al.* 2013) and R-V measurements of the stars have put upper limits of < 10 $M_{Jup}$ on any close-in companion masses within 50 AU of their primaries (Galland *et al.* 2010, priv commun.).

## 2.3  CXO Spectroscopy. 

From the 1370 detected events, we produced a total HD 113766A+B spectrum, extracted over both sources with an r=20 pixel circular aperture. The combined spectrum is shown in Figure 3a, and is rather soft. We estimate a total HD 113766 A+B system luminosity of $L_x$ = 2.2 x $10^{29}$ erg $s^{-1}$. We also extracted separate spectra for the "West" and "East" sources, each over an r=1.6 pixel circular aperture (Fig 3a). After allowing for 30 counts from the brighter Western source in the fainter Eastern source's aperture, this provided a reasonable separation of the two source photon populations, with 1200 counts for the western source and 93 counts for the eastern source in the 0.3-2.0 keV range, implying luminosities $L_x$ ~ 2.0 x $10^{29}$ erg $s^{-1}$ for the Western source (HD 113766B) and $L_x$ ~ 1.6 x $10^{28}$ erg $s^{-1}$ for the Eastern source (HD 113766A)

We show model fits to the spectrum of the brighter HD 113766B source (Figure 3b). Assuming solar metallicity (following Fe/H = -0.01 from Nordström *et al.* 2004), the spectrum can be fit by an APEC emission model spectrum for collisionally-ionized diffuse gas calculated using the ATOMDB code v2.0.1. The best-fit model appears to be very soft, with an APEC temperature of 0.50 ± 0.06 keV (or 6.7 ± 0.8 x $10^6$ K), and a two-temperature model was not required to fit the data. While 4 to 6 times hotter than the effective coronal temperature of the 4.5 Gyr old Sun (~0.10 keV), it is about the temperature found for other 10 - 20 Myr solar type stars of similar $L_x$ (Suchkov *et al.* 2003, Telleschi *et al.* 2005). The shape of the spectrum, peaking at ~0.8 keV, is also similar to the examples shown in Telleschi *et al.* 2005 in their X-ray spectral survey of young solar type stars. No strong emission lines above the background are obvious, but this is likely an effect of the coarse energy resolution of ~50 eV (1σ) of the ACIS-S CCD coupled with the low number of total counts. The total flux detected is 1.4 x $10^{-13}$ ergs cm$^{-2}$ s$^{-1}$, from 0.3 to 2.0 keV. The formal derived hydrogen upper limits of $N_H < 10^{20}$ cm$^{-2}$ for the APEC models are consistent with the ~5 x $10^{19}$ cm$^{-2}$ expected for a stellar source ~123 pc distant separated by an intervening interstellar H density of 0.1 particles /cm$^3$, suggesting that there is little in-system H absorption. A small hydrogen column is also consistent with there being little extinction towards the binary, as evidenced by comparing the observed B-V for the system of 7.91-7.56 = 0.35 to the predicted intrinsic B-V = 0.43 (Pecaut *et al.* 2012).

With only 93 total counts, the total number of events obtained for the fainter HD 113766A "East" source is so small that we cannot used binned statistics for spectral fitting. Instead, we use a likelihood-based method (Cash statistic) which is appropriate for Poisson distributed data. Forcing the $N_H$ column to be the same as used in our HD 113766B modeling, i.e. $N_H < 10^{20}$ cm$^{-2}$, we find that the best fit is a stellar source with a very cool APEC temperature of ~0.38 keV or 5.6 x $10^6$ K, about 75% the temperature of the B source, and a total flux of 0.14 x $10^{-13}$ ergs cm$^{-2}$ s$^{-1}$, corresponding to an X-ray luminosity $L_x = 1.6$ x $10^{28}$ erg s$^{-1}$ from 0.3 to 2.0 keV. As the low-count likelihood fitting method does not produce a formal $\chi^2$ statistic we cannot derive formal confidence limits, but via forward modeling Figure 3 shows that temperatures in the range 0.32 - 0.44 keV (or 4.3 - 5.9 x $10^6$ K) produce reasonable fits to the spectra.

## 3. Analysis.

In this section we discuss the first-order implications of our Chandra observations of HD 113766 in comparison to x-ray observations of other stars. In the next section, §4, we will discuss their connections to the bigger picture of debris disk evolution.

**3.1 Luminosity Results.** Using Chandra, we have found an HD 113766 system with *Chandra* count rate and luminosity very close to the published RASS values. The observed x-ray emission can be accounted for by two stellar coronal sources located at the positions of HD 113766A and HD 113766B. HD 113766A is more than 12x fainter in the X-ray, consistent with the fact that it is the earlier of the two stars (Pecaut *et al.* 2012 classified HD113766A as F1-F3V, and Chen *et al.* 2011 have classified HD 113766B as F5V - F7V). The fact that HD 113766A is relatively X-ray faint at ~2 x $10^{28}$ erg s$^{-1}$, at the lowest end of the *Hipparcos*-ROSAT survey luminosity range (Suchkov *et al.* 2003) is also consistent with it being a very early F-star, as is comparison of our Chandra HD113766A results to those found by Feigelson *et al.* (2006) for 51 Eri, a 23 ± 3 Myr (Bell *et al.* 2015) F0V star with kT = 0.2 and $L_x$ = 1.4 x $10^{28}$ erg s$^{-1}$.

Similarly, the $L_x$ = 2 x $10^{29}$ erg s$^{-1}$ Chandra luminosity we find for HD113766B is consistent with the $L_x$ ~ $10^{29}$ erg s$^{-1}$ luminosity reported for SAO 206462 by Müller *et al.* (2011), a young Herbig F8V star with a reported 3.9h rotation period. An X-ray luminosity of $10^{29}$ to $10^{30}$ erg s$^{-1}$ is high for the average main sequence star, but from the Kepler study of the rotation rates of stars (Meibom *et al.* 2011), as well as previous Hyades and Pleiades measurements (Stern *et al.* 1995; Stauffer *et al.* 1994; Güdel 2004) it is plausible for the very young HD 113766 F-stars to be brighter in the X-ray by 1-2 orders of magnitude than their mature main sequence F-star counterparts.

**3.2 Imaging Results.** Our *Chandra* data has, for the first time, resolved the two stars of HD 113766 in X-rays (Fig. 2a). Examining in detail our deconvolved *Chandra* X-ray imagery (Fig. 2b), we do not see any evidence for extended X-ray emission, or emission produced by anything but two point sources to the resolution limit of our mapping (~0.2", or 25 AU at 123 pc distance).

The main unusual finding we have for this system is its brightness asymmetry. One of the main questions raised by this was the question of outbursts – i.e., was it possible that HD 113766B was flaring during the *Chandra* observations? Three lines of inquiry suggest that it was not: (a) the achieved count rate for the *Chandra* observations was within 10% of the rate estimated from the 1989 RASS counts, suggesting the system X-ray luminosity had been stable over 21 years; (b) the time series of photons detected in our *Chandra* observations show no variability over and above a possible ~2.9 hr amplitude periodicity in the West source; and (c) the temperature of the HD 113766B X-ray spectrum is low compared to the typical 10 - 100 MK flare temperatures in stellar coronae (Feldman *et al.* 1995, Kashyap *et al.* 2002, Shibata & Yokoyama 2002).

Given the incredibly dusty nature of the HD 113766A-system and the lack of any dust signature in the HD 113766B-system, coupled with the binary pair's coeval age of 10-16 Myr and the observed asymmetry in X-ray production, a natural inference from our results is that strong X-ray and stellar wind emission destroy and/or accelerate the removal of dust from a stellar system. Support for the latter finding comes from estimates of highly reduced dust lifetimes due to stellar wind drag (Chen *et al.* 2005, 2006, 2011), and by the anti-correlation between the lifetime of circumstellar dust and primary star X-ray luminosity found in TW Hya by Kastner *et al.* (2004, 2016). We will discuss this in more detail in Section 4.

**3.3 Spectral Results.** Spectrally, the observed HD 113766B X-ray emission is soft with $kT \sim 0.50$ keV. This temperature is a factor of a few lower than a typical young G0 (c.f. Preibisch et al 2005). This is not expected if the source of the flux is an $\alpha$–$\Omega$ dynamo, especially for a system in which there is evidence of rapid stellar rotation (Fig 1), unless the shear at the base of the convection zone is especially weak. Other possible effects on the observed spectrum, e.g. a low coronal electron density or absorption by a large column of nearby H or absorption through the disk of circumstellar material would reduce the luminosity dramatically as well as selectively absorb lower energy photons, is clearly not consistent with the observed ACIS-S spectrum for HD 113766B. The possibility that the soft spectrum can be attributed to accretion is remote, as (a) Chen *et al.* (2011) did not report any detectable H-alpha emission in their Magellan/MIKE R~50,000 spectra for HD 113766, and (b) because the diskless star of the two in the binary is dominating the X-ray flux. Further, R~10,000 NIR spectroscopy of the HD 113766A+B system using the SPeX instrument at the NASA/IRTF 3m in 2011 obtained by our group did not detect any CO or HI Brackett γ line emission which would be expected in the case of ongoing accretion (Connelly & Greene 2010, 2014; Lisse *et al.* 2012, 2015). Related NIR measures of outflow activity in our SPeX spectrum, using FeII, HeI, and $H_2$ lines (Connelly and Greene 2014) are absent, arguing against outflow shocks as a soft X-ray source as well.

**4. Discussion.** F-stars are intriguing objects in the X-ray. At the early end of their type, they are highly radiative and minimal X-ray emitters. At the other end of their type, they are highly convective and are strong X-ray emitters. Myr-old F-stars can be expected to be fast rotators, and more convective and X-ray active than their main sequence cousins. The HD 113766 system has been alternately described as a 10-16 Myr old F4/F6 or F3/F5 spectroscopic binary by most observers using optical photometry (Holden 1975, 1976; Houk 1978, Olsen & Perry 1984, Mannings & Barlow 1998, Hauck & Mermilliod 1998, De Zeeuw *et al.* 1999, Hoogerwerf *et al.* 2000, Fabricius & Makarov 2000, Madsen *et al.* 2002, Sartori *et al.* 2003, Hodge *et al.* 2004, Nordstroem *et al.* 2004, Chen *et al.* 2005, Rhee *et al.* 2008), so we would have naively expected both stars to be rapidly rotating, highly convective, and X-ray bright. The fact that our East (A) source is 12x fainter in the X-ray is consistent with it being the earlier of the two stars. The fact that HD 113766A is X-ray faint in the absolute sense at ~$10^{28}$ erg $s^{-1}$, at the lowest end of the *Hipparcos*-ROSAT survey (Suchkov *et al.* 2003) suggests that it is even earlier than F3V to F4V, closer to F2V, and the East component is a late, much more x-ray active F6V star, very much in agreement with Pecaut *et al.*'s (2012) recent optical spectroscopic re-assessment of the system.

It also suggests that any stellar wind and high energy stellar radiation effects on circumstellar material and objects are much more important in the HD 113766B system than the HD 113766A system, although the models of Ciesla and Sandford (2012) argue that high energy irradiation can drive important levels of organic synthesis in any water and organic-rich dust present in either system – e.g., as has been reported in HD 113766A by Lisse *et al.* (2008). If the solar system's history is a guide, this irradiated material is likely to be astrobiologically important, as the larger pieces of it can be re-incorporated into asteroids and planetesimals aggregating during the terrestrial planet building era (which occurred from age = 10 to 100 Myr in our system). Many of these bodies will later accrete onto the terrestrial planets present in the system during the equivalent of our solar system's Late Veneer and Late Heavy Bombardment eras (Bottke *et al.* 2010, Raymond *et al.* 2013).

**4.1 Dust – $L_x$ Anti-Correlation.** We note with interest the highly dusty nature of the X-ray faint HD 113766A-system and the dust poor nature of the X-ray bright HD 113766B-system. The X-ray observations of TW Hya stars by Kastner *et al.* (2004, 20016), and the fact that our own Sun is X-ray faint and planet rich vs. the Kepler G-star average (Basri *et al.* 2010, 2011) suggests the distinct possibility of a causal connection between circumstellar dust excesses and low X-ray luminosities. I.e., it would seem that the asymmetric X-ray flux found for HD 113766A vs. HD 113766B by *Chandra* for the two coeval, similarly sized F-stars in this system requires a nature or nurture explanation - do dusty disks somehow diminish the observed (but corrected for dust absorption) X-ray flux (nature), or do disks thrive longer in low X-ray and stellar wind flux systems (nurture)?

To address the question of potential X-ray obscuration, it is important to consider the possibility that the circumstellar dust in HD 113766A is strongly attenuating the primary's coronal emission via absorption and scattering. A simple calculation of $L_{IR}/L_{bol}$ (Lisse *et al.* 2008 and references therein) shows that while the HD 113766A disk is very massive, it intercepts and scatters at most $\sim 10^{-3}$ of the bolometric (mostly optical) flux of the star. Assuming normal X-ray scattering cross sections (Morrison & McCammon 1983) and a thin disk geometry, it is difficult to see how the HD 113766A primary could be emitting roughly the same $L_x \sim 2 \times 10^{29}$ erg s$^{-1}$ as HD 113766B, only to have 90% of this flux absorbed by intervening circumstellar dust. Some enhancement of the observed extinction might be possible if the HD 113766A dust belts (Lisse *et al.* 2008) were in a fortuitous edge-on viewing situation, à la the Beta Pic system. However, infrared imaging of the system does not argue for an unusual edge-on dust disk viewing geometry (Meyer *et al.* 2001), and the Spitzer IRS spectrum of the dust shows strong emission features, arguing against optically thick dust belts. Further, the existence of soft ( < 0.5 keV) X-rays from HD 113766A (Fig 3) indicates a very small amount of Hydrogen gas along the line of sight (< $10^{20}$/cm$^2$) and thus a very small Av < 1, even for gas to dust ratios as small as 25:1. Thus dust obscuration of the HD 113766A X-rays is an unlikely reason for the observed X-ray asymmetry, and we again recover the reason for the star's x-ray luminosity asymmetry is due to their differing stellar types.

On the other hand, previous authors have noted a possible dusty disk–X-ray luminosity anti-correlation due to the effects of a star's high energy irradiation on material in orbit around it. *Chandra* observations of HD 98800, a quadruple system in the 10 Myr old TW Hydrae association, have revealed that the X-ray flux of the dusty binary system HD 98800B is 4x fainter than its dustless companion HD 98800A (Kastner *et al.* 2004). Ground-based searches for new 10 and 20 μm excesses around proper-motion and X-ray selected K- and M-type members of the 10 Myr old TW Hydrae association (Weinberger *et al.* 2004) and around proper-motion, X-ray, and lithium-selected F-, G-, K-, and M-type members of the 30 Myr old Tucana-Horologium association (Mamajek *et al.* 2004) have been relatively unsuccessful. Recent work by Kastner *et al.* (2016) targeting the TW Hya M-star population has found an anti-correlation between a star's photospheric temperature /Lx and the amount of IR excess flux arising from circumstellar material surrounding it.

Chen *et al.* (2005) summarized the possibilities when they wrote in examining their large sample of stars from the young, nearby Sco-Cen stars forming region: "…Contrary to expectation, the infrared luminosity appears anti-correlated with X-ray luminosity, except for 30% of the objects that possess neither a ROSAT flux nor a MIPS 24 μm excess. The anti-correlation can be naturally explained if stellar wind drag effectively removes dust grains around young stars with high X-ray coronal activity….". Chen *et al.* (2011) reiterated this possible correlation, stating in the conclusions of their updated ~400 star Sco-Cen debris disk survey that they have found a "weak anti-correlation" between the ROSAT flux and the mid-IR flux from a system (the main problems with finding a stronger correlation were stated as the limiting sensitivities of the ROSAT survey and the derived stellar mass loss rates).

**4.2 The Effects of Stellar Wind Drag.** Chen *et al.* (2005, 2006, 2011) also noted the unusual nature of the HD 113766 system and its ROSAT detection, suggesting the important role that stellar wind drag could have in determining the lifetime of dust in a circumstellar debris disk versus infall onto the primary star. Specifically, they argued that the effects of a dense stellar wind on orbiting dust are similar to those of photons causing Poynting-Robertson drag, with the total contributions of the two mechanisms to the infall velocity $v_{infall}$ going as $v_{P-R}$ * $(1 + c^2 [dM/dt]_{wind} /L_*)$. For the Sun, we have $[dM/dt]_{wind, Sun}$ = 2.0 x $10^{12}$ g s$^{-1}$ and $L_*$ = 3.9 x $10^{33}$ erg s$^{-1}$, implying $v_{infall} \sim 1.47$ $v_{P-R}$.

Photospheric X-ray emission is connected to stellar wind flow, as x-ray emission comes from the breaking of magnetic field lines, and the now open field lines guide energetic plasma away from the star, creating a stellar wind (Osten & Wolk 2015). This allows us to estimate the stellar wind mass loss rate using $[dM/dt]_{wind}$ = C * $4\pi R^2$ $F_{x,*}^{1.3}$, where C is a constant, $R_*$ is the stellar radius, and $F_{x,*}$ is the X-ray flux per unit stellar surface area (Wood *et al.* 2002, 2005, 2014). The constant C is determined by scaling from the solar result with $F_{x,Sun}$ = 3.7 x $10^4$ erg cm$^{-2}$ s$^{-1}$ (Mamajek *et al.* 2002; Wood *et al.* 2002, 2005).

Assuming $R_{HD113766A} \sim R_{HD113766B}$ = 1.35 $R_{sun}$ (Lisse *et al.* 2008 and references therein), with our new *Chandra* results for the binary's X-ray fluxes we have $F_{x, HD113766A}$ = 1.4 x $10^5$ erg/cm$^2$/sec and $F_{x, HD113766B}$ = 1.8 x $10^6$ erg cm$^{-2}$ s$^{-1}$ (as compared to the $F_{x,*}$ = 9.2 x $10^5$ erg cm$^{-2}$ s$^{-1}$ for both stars of the binary estimated by symmetrically assigning the RASS flux of the system; Chen *et al.* 2011). This implies $[dM/dt]_{wind, HD113766A}$ = 2.3 x $10^{13}$ g s$^{-1}$ and $[dM/dt]_{wind, HD113766B}$ = 5.7 x $10^{14}$ g s$^{-1}$. With $L_{HD113766A}$ = 4.4 $L_{Sun}$ for the F2V A-member (Lisse *et al.* 2008 and references therein; Chen *et al.* 2011), we find $v_{infall}$ = (2.2 ± 0.24) $v_{P-R}$ for HD 113766A. Assuming $L_{HD113766B} \sim$ 2.3 $L_{Sun}$ for the F6V B-member (as the model ratio of F6V luminosity to F2V luminosity is ~0.52), we have $v_{infall}$ = (58 ± 3.3) $v_{P-R}$ for HD 113766B.

We thus find, from our *Chandra* results, that the total drag effects for HD 113766B are ~26 times larger as for HD 113766A, and that stellar wind drag effects easily dominate the dynamical repulsive forces for HD 113766B. For HD113766A, like our solar system, stellar wind drag effects are about equal in effect vs. radiative drag forces. However, since $v_{P-R}$ ~ $L_*$ (Burns *et al.* 1979), dust in the HD113766A system still falls onto the primary about 6.6 times faster than in our solar system; but this is still rather slow compared to the 91 times faster infall rates experienced around HD113766B. In absolute terms, we have $t_{infall} = t_{infall,solar-system}$ * $(v_{infall,solar-system}/v_{infall,*})$ ~ 400 * $D^2$ / (0.2/$r_{dust}$ (um)) yrs * $(v_{infall,solar-system}/v_{infall,*})$ (where D = the distance from the dust particle to the central star and $r_{dust}$ is the dust particle radius; Burns *et al.* 1979). For the warm dust located at ~1.8 AU from the HD 133766A primary (Lisse *et al.* 2008), assuming it consists of a population of dust of which the largest and longest lived grains initially present are the abundant ~1 mm sized dust grains seen in solar system chondrules and comet trails common to the early solar system, this implies an infall time of 1.5 x $10^6$ yrs for HD 113766A, or a disk clearing time on the order of 1/10th the primary's estimated total age. This suggests that the stability of a dense circumstellar dust disk, massing as least as much as Mars and created by currently ongoing intense asteroidal grinding or planetary accretion (Lisse *et al.* 2008; Olafsson *et al.* 2013), is quite plausible. Doing the same calculation for HD 113766B, we find a much shorter infall and disk clearing time of ~0.9 x $10^5$ yrs, less than 1% of the primary's estimated age, and it is not surprising that this system has quickly cleared out any dust created by collisions or impacts onto growing planetary embryos.

In summary, our *Chandra* HD 113766 binary X-ray results, taken together with those for the HD 98800 binary and other singleton debris disks (Kastner *et al.* 2004, 2016; Chen *et al.* 2005, 2011; Glauser *et al.* 2009), suggest that a requirement for circumstellar dust longevity is the lack of a strong primary stellar wind. If this is correct, then we can also expect that on the average young, early F-type stars should have a higher frequency of circumstellar dust disks than young, late type F-stars as the X-ray luminosity and stellar wind activity increase across the class. It is also interesting to think that a relatively low X-ray and solar wind flux may have been the operative case in the young solar system – as suggested by the low activity rate for the Sun versus G-stars found in the Kepler sample (Basri *et al.* 2010, 2011).

**5. Conclusions.** We have used the *Chandra* X-ray Observatory (CXO) to obtain imaging spectroscopy of the close (1.4″, or 170 AU separation), coeval (10-16 Myr old) F-star binary HD 113766 over 38 ksec. All 3 *Chandra* low energy X-ray measures of this object – imaging, photometry, and spectroscopy – show a system with 2 detectable sources separated along an E-W line by ~1.4″, with the W source approximately 10 times as bright as the E source. The emission spectrum of each object is well fit by an APEC coronal emission model, although the emission appears to be rather soft for such young stars, $kT = 0.30 – 0.50$ keV, leading us to suspect that the coronal magnetic fields are weak in these F stars. We find asymmetric X-ray emission from the two stellar sources, with the emission from the easternmost, the IR- extended primary object HD 113766A, only ~10% that of the western star HD 113766B. There is no evidence for a 3rd member of the HD113766 with mass greater than 0.1 $M_{Sun}$ within 2' of the AB pair. The X-ray emission from the HD 113766B stronger source may vary with a $2.8 \pm 0.15$ hr period. For both stars, the strength of the X-ray emission varies inversely with the excess infrared flux from circumstellar material. Stellar wind drag models corresponding to the $L_x \sim 2$ x $10^{29}$ erg s$^{-1}$ argue for a dust lifetime around HD 113766B of ~90,000 years, suggesting that HD 113766B efficiently clears any secondary dust out of its system, whereas HD 113766A, with $L_x \sim 2$ x $10^{28}$ erg s$^{-1}$ (12 times fainter than B) and dust lifetime $> 1.5$ x $10^6$ years, could have created the dust seen today anytime within the last Myr. A similar situation has been found for a few other young debris disks, most notably HD 98800 by Kastner *et al.* (2004). Over the course of 1 Myr, the HD 113766A X-ray emission and stellar wind irradiation is high enough to drive important levels of organic synthesis in the orbiting circumstellar material, which is rich in water and carbonaceous materials (Lisse *et al.* 2008; Cielsa and Sandford 2012).

**6. Acknowledgements.** The authors would like to thank J. Kastner, E. Mamajek, and J. Raymond for many useful discussions concerning X-ray emission from young stellar sources. C.M. Lisse gratefully acknowledges support for this work provided by the National Aeronautics and Space Administration through *Chandra* Award Number GO1-12028X issued by the *Chandra* X-ray Observatory Center (CXC). The CXC is operated by the Smithsonian Astrophysical Observatory for and on behalf of the National Aeronautics Space Administration under contract NAS8-03060.

## 8. Figures

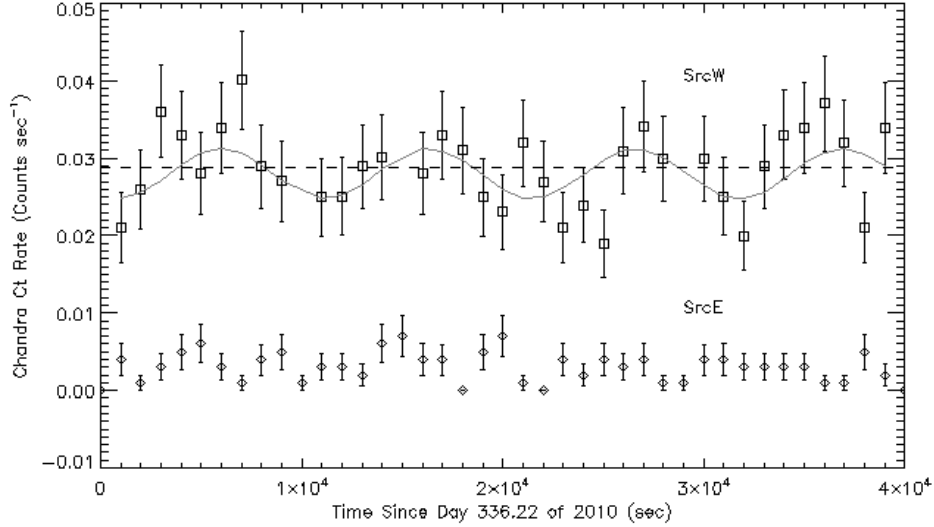

**Figure 1 – (a)** *Chandra* **ACIS-S photometry of the two X-ray sources detected in the HD 113766 system.** All error bars and error estimates are 2σ. The background count rate in the *Chandra* PSF, as measured far off-source, has been measured with level 0.01 +/- 0.002 cps and removed from these curves. The western source, labeled "SrcW", and identified with HD 113766B, demonstrates a flux of 0.029 +/- 0.005 cps (squares), with a possible modulation with sinusoidal periodicity of 2.8 (± 0.15) hours and amplitude 0.0033 (± 10%) cps (grey curve) ("possible" since a constant flux model [dashed line] also has $\chi^2$ value within the 95% confidence limits for our 36 dof lighturve). The eastern source, labeled as "SrcE" and identified with HD 113766A (diamonds), is roughly 10 times fainter at 0.0037 ± 0.002 cps. The detection of SrcE is at too low a significance to determine any modulation in the A source. **(b)** Power spectrum of the HD 113766B *Chandra* lightcurve (black) compared to that of the background (green), with the location of the possible 2.8 hr periodicity and its n=2 and n= 4 harmonics marked by the red dashed lines.

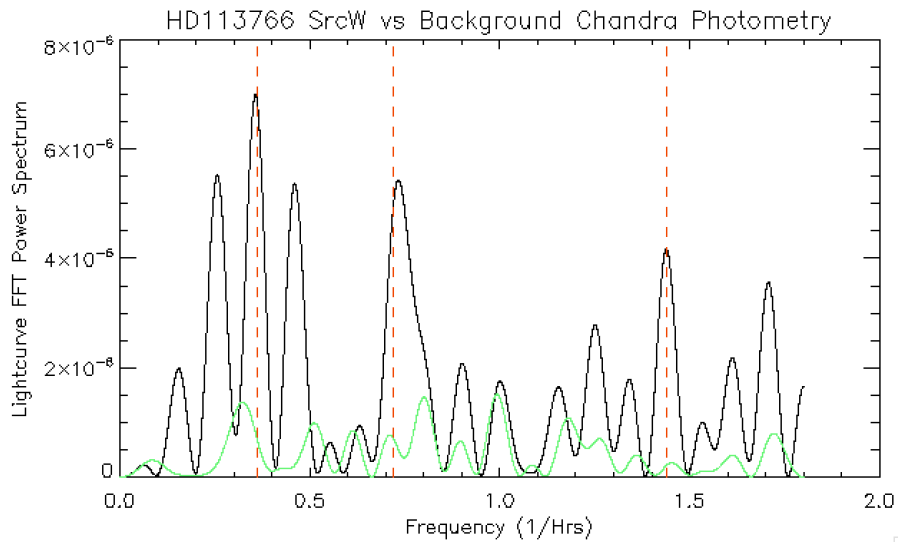

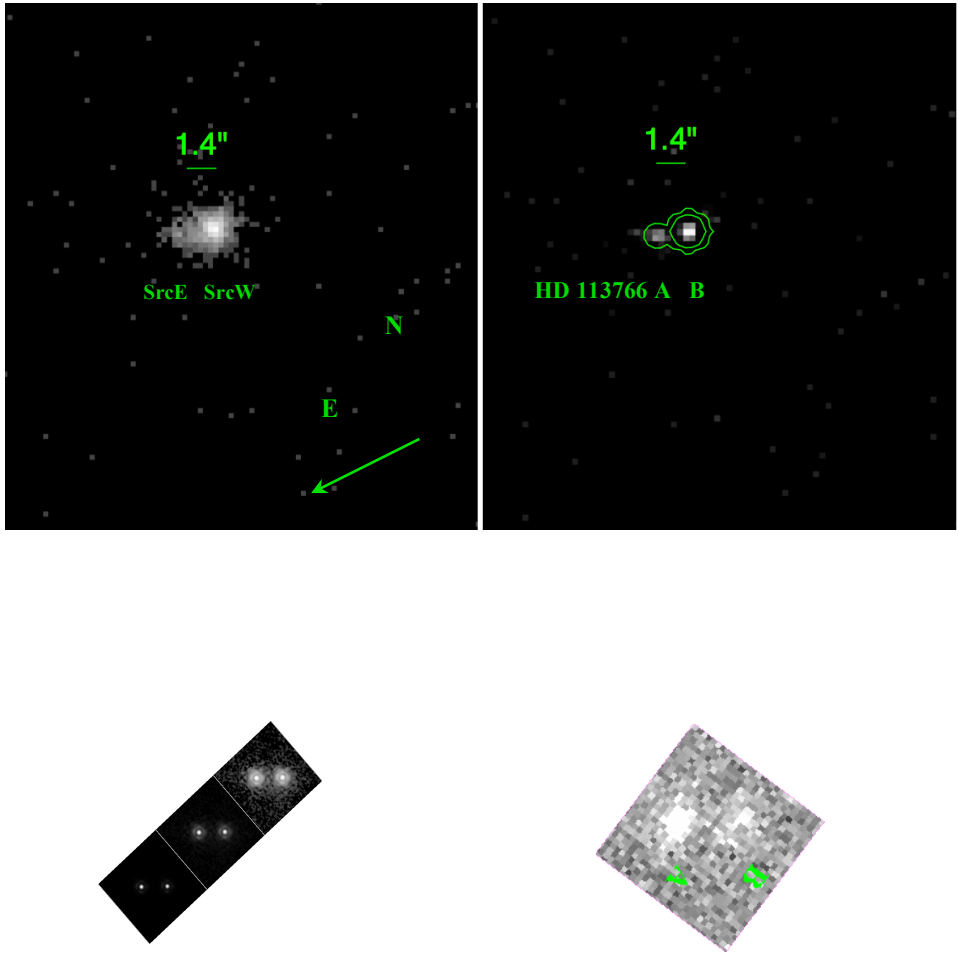

**Figure 2 – *Chandra* ACIS-S imagery of the HD 113766 system. (a, Upper Left)** Raw imagery of the system, with all detected photons mapped onto the sky plane. Here E is to the Left and North is up, implying that the Western stellar member of the binary is much more X-ray bright than the Eastern member. **(b, Upper Right)** Deconvolved map of the *Chandra* observations, showing the source separation with 0.25" pixels. **(c, Lower Left)** HST/NICMOS 1.1 μm image of the system (After Meyer *et al.* 2001). **(d, Lower Right)** : Magellan 11 μm image of the system (After Meyer *et al.* 2001, Smith *et al.* 2012).

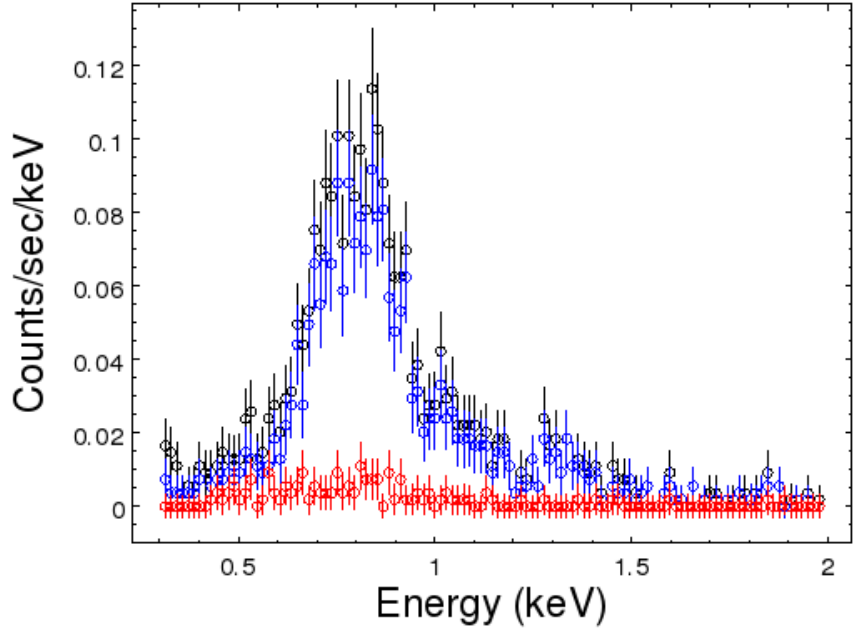

**Figure 3 – *Chandra* ACIS-S spectroscopy of the HD 113766 system.** (Above) Spectrum of the combined E+W counts (black circles), bright Western source (HD 113766B, blue circles), and fainter Eastern source (HD 113766A, red circles). (Below Left) Plot of calibrated ACIS-S X-ray spectrum and APEC coronal model fits to the West source counts (solid colored lines - orange = 0.32 keV, light green = 0.44 keV, dark blue = 0.56 keV, and olive = 0.68 keV). The 0.44 and 0.56 keV models both fit the data equally well. (Below right) Plot of calibrated ACIS-S X-ray spectrum and APEC coronal model fits to the East source counts (solid colored lines - red = 0.20 keV, orange = 0.32 keV, light green = 0.44 keV, dark blue = 0.56 keV, and olive = 0.68 keV). The 0.32 and 0.44 keV models fit the data the best.

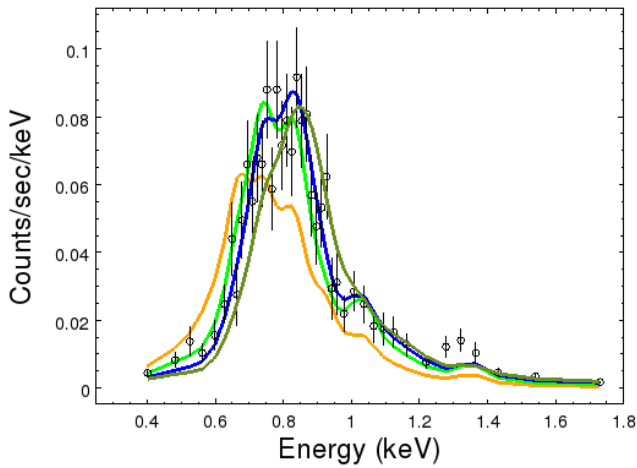 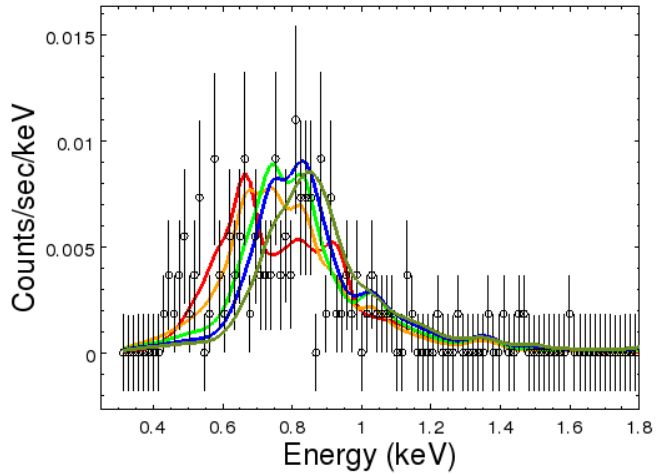